\documentclass[12pt]{article}
\usepackage[dvips]{graphics}
\usepackage{amsmath}
\usepackage{epic}

\makeatletter
\renewcommand{\@makecaption}[2]{
  \vskip\abovecaptionskip
  \sbox\@tempboxa{\small\sf #1: #2}%
  \ifdim \wd\@tempboxa >\hsize
  \small\sf #1: #2\par
  \else
    \global \@minipagefalse
    \hb@xt@\hsize{\hfil\box\@tempboxa\hfil}%
  \fi
  \vskip\belowcaptionskip}
\makeatother

\newcommand{\preprint}[1]{\rule{0pt}{8pt} \scriptsize #1}

\numberwithin{equation}{section}

\providecommand{\abs}[1]{\lvert#1\rvert}

\date{}
\title{(De)Constructing Dimensions}
\author{Nima Arkani-Hamed\thanks{{\tt
      arkani@bose.harvard.edu}\hfil\break
    Permanent address: \small\sl Department of Physics, UC Berkeley,
    Berkeley, CA 94720}, \ Andrew G. Cohen\thanks{{\tt
      cohen@andy.bu.edu}\hfil\break
    Permanent address: \small\sl Physics Department, Boston
    University, Boston, MA 02215} \ and\ 
  Howard Georgi\thanks{\tt georgi@physics.harvard.edu}\\ \\
  \small\sl Lyman Laboratory of Physics \\
  \small\sl Harvard University \\
  \small\sl Cambridge, MA 02138
}

\newsavebox{\moose}
\sbox{\moose}{%
\begin{picture}(0,0)
  \thicklines
  \put(-60,0){\circle{20}}
  \put(60,0){\circle{20}}
  \put(-50,0){\line(1,0){40}}
  \put(0,0){\circle{20}}
  \put(10,0){\line(1,0){40}}
  \put(-25,0){\vector(1,0){0}}
  \put(35,0){\vector(1,0){0}}
\end{picture}}
\newsavebox{\cmoose}
\sbox{\cmoose}{%
\begin{picture}(0,0)
  \thicklines
  \put(-60,0){\circle{20}}
  \put(60,0){\circle{20}}
  \dashline{6}(-50,0)(50,0)
  \put(0,0){\vector(1,0){0}}
\end{picture}}

\begin{document}
\begin{titlepage}
  \maketitle
  \begin{picture}(0,0)
    \put(400,200){\shortstack{
        \preprint HUTP-01/A015\\
        \preprint BUHEP-01-05\\
        \preprint LBNL-47676\\
        \preprint UCB-PTH-01/11\\
        \rule{0pt}{8pt} }}
  \end{picture}
  
  \begin{abstract}
    We construct renormalizable, asymptotically free, four dimensional
    gauge theories that dynamically generate a fifth dimension.
  \end{abstract}
  \thispagestyle{empty} \setcounter{page}{0}
\end{titlepage}

\section{Introduction}
\label{sec:introduction}

The world is apparently four-dimensional. But it is possible that at
distances shorter than those yet probed the Universe may best be
described by a theory with more than the conventional one time and
three space coordinates. A simple model of such extra dimensions is a
theory of fields living on a spacetime with four extended dimensions,
plus one or more additional compact dimensions. At distances large
compared to the size of these compact dimensions, such a theory
appears four dimensional: gauge forces fall off like the square of the
distance, free energies of massless degrees of freedom scale like the
fourth power of the temperature, \textit{etc.} At energies
corresponding to the inverse compactification size, Kaluza-Klein
excitations appear with a spectrum dictated by the detailed nature of
the compact space.  At energies much higher than this scale, the extra
dimensions become manifest: physics at distances small compared to the
compactification size is insensitive to the compactification, and the
theory appears higher dimensional.

Unfortunately these higher dimensional field theories have
dimensionful couplings and therefore require a cut-off. As energies
approach this cut-off, physics depends sensitively on the cut-off
procedure, typically becoming strongly coupled. This makes it
difficult to address what happens at energies above the cut-off.
Indeed, quantum gravity in four dimensions is challenging for similar
reasons.  Nevertheless, a few UV completions of higher-dimensional
field theories have been suggested, each realizing the
higher-dimensional theory as the low energy limit of some more
fundamental theory with a sensible high energy behavior. One
possibility is that the cut-off of the higher dimensional field theory
coincides with the fundamental Planck scale, where gravity also
becomes strong. In itself this does not allow us to say anything about
the behavior of the theory at energies above the cut-off since
super-Planckian quantum gravity is poorly understood. Moreover, the UV
difficulties with higher-dimensional field theories are unrelated to
gravity, and it is therefore interesting to search for UV completions
of higher-dimensional field theories where gravity is completely
decoupled. Some examples of this kind have emerged in
non-gravitational subsectors of superstring theory, including (0,2)
super-conformal theories, little string theories and open-membrane
theories of various kinds~\cite[for
example]{Seiberg:1998ax,Gopakumar:2000ep}. Unfortunately these
theories are strongly coupled and typically difficult to understand.
Furthermore, they can not be defined in more than six dimensions, and
seem to rely on unbroken supersymmetry in an essential way.

In this paper we describe a new way of UV completing
higher-dimensional field theories. Instead of starting with extra
dimensions, we \emph{build} them.  In an inversion of the usual
picture, these models are \emph{four-dimensional} at very high
energies. They are renormalizable and in most cases even
asymptotically free.  Extra dimensions emerge \emph{dynamically} at
low energies, in a simple and calculable way. This allows us to study
many mysterious features of higher-dimensional field theories on a
firm footing, without worrying about the unknown physics of the UV
cutoff.  Even more important, our construction of extra dimensions
puts higher-dimensional physics into a broader context, and serves as
a departure point for exploring more radical and even more interesting
new possibilities.

\section{$SU(n) \times SU(m)$ moose}
\label{sec:moose}

Our example field theories, all of which will be four dimensional,
contain gauge fields and fermions, and are conveniently summarized in
a pictorial representation, referred to variously as
``moose''~\cite[for example]{Georgi:1986hf} or ``quiver''~\cite{Douglas:1996sw}
diagrams.  In such diagrams gauge groups are represented by open
circles, and fermions by single directed lines attached to these
circles. A line directed away from a circle corresponds to a set of
Weyl fermions transforming as the fundamental representation of the
gauge group, while a line directed toward a circle corresponds to a
set of Weyl fermions transforming as the complex conjugate of the
fundamental representation. The moose diagram we will consider is the
$N$-sided polygon
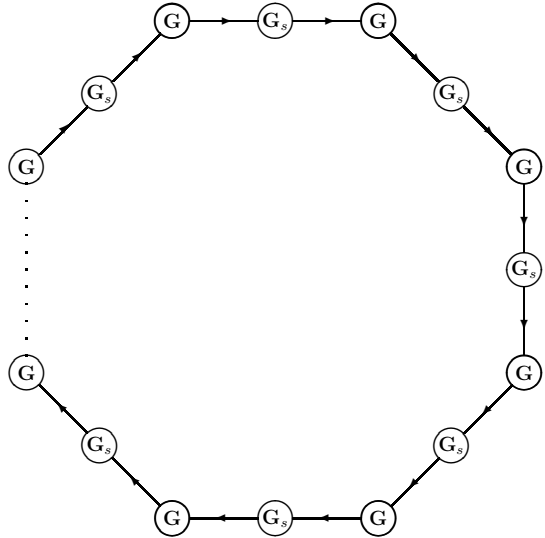
\begin{figure}[htb]
  \centering
  \scalebox{.65}{
    \begin{picture}(0,325)(0,-144.8)
      \thicklines
      \put(-102.4,102.4){\rotatebox{45}{\usebox\moose}}
      \put(0,144.8){\usebox\moose}
      \put(102.4,102.4){\rotatebox{-45}{\usebox\moose}}
      \put(144.8,0){\rotatebox{-90}{\usebox\moose}}
      \put(102.4,-102.4){\rotatebox{-135}{\usebox\moose}}
      \put(0,-144.8){\rotatebox{-180}{\usebox\moose}}
      \put(-102.4,-102.4){\rotatebox{-225}{\usebox\moose}}
      \put(-144.8,0){\rotatebox{270}{%
          \makebox(0,0){\dottedline{10}(-50,0)(50,0)}}}
      \put(-60,144.8){\makebox(0,0){\bf G}}
      \put(60,144.8){\makebox(0,0){\bf G}}
      \put(60,-144.8){\makebox(0,0){\bf G}}
      \put(-60,-144.8){\makebox(0,0){\bf G}}
      \put(144.8,60){\makebox(0,0){\bf G}}
      \put(144.8,-60){\makebox(0,0){\bf G}}
      \put(-144.8,60){\makebox(0,0){\bf G}}
      \put(-144.8,-60){\makebox(0,0){\bf G}}
      \put(0,144.8){\makebox(0,0){$\mbox{\bf G}_s$}}
      \put(0,-144.8){\makebox(0,0){$\mbox{\bf G}_s$}}
      \put(144.8,0){\makebox(0,0){$\mbox{\bf G}_s$}}
      \put(102.4,102.4){\makebox(0,0){$\mbox{\bf G}_s$}}
      \put(102.4,-102.4){\makebox(0,0){$\mbox{\bf G}_s$}}
      \put(-102.4,102.4){\makebox(0,0){$\mbox{\bf G}_s$}}
      \put(-102.4,-102.4){\makebox(0,0){$\mbox{\bf G}_s$}}
    \end{picture}}
  \caption{A moose diagram.}
  \label{fig:moose1}
\end{figure}
representing a field theory with a $\text{\bf G}^N \times \text{\bf
  G}_s^N$ gauge group and fermions transforming bi-linearly under
``nearest-neighbor'' pairs of gauge transformations.

For definiteness we will take $\text{\bf G}=SU(m)$ and $\text{\bf G}_s
= SU(n)$. We will impose a cyclic symmetry to keep all $SU(m)$ gauge
couplings equal to a common value $g$, and all $SU(n)$ gauge couplings
equal to $g_s$. By dimensional transmutation we may equally well
describe this theory by two corresponding dimensionful parameters,
$\Lambda$ and $\Lambda_s$.  Each side of this polygon describes two
types of fermions transforming under the three gauge groups associated
with this side, $SU_i(m)\times SU_i(n)\times SU_{i+1}(m)$:
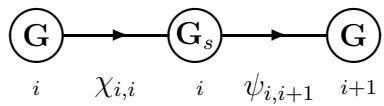
\begin{center}
  \begin{picture}(0,40)
    \put(0,20){\usebox\moose}
    \multiput(-60,20)(120,0){2}{\makebox(0,0){\bf G}}
    \put(0,20){\makebox(0,0){$\mbox{\bf G}_s$}}
    \multiputlist(-60,0)(60,0){$\scriptstyle i$, $\scriptstyle i$,
      $\scriptstyle i+1$}
    \multiputlist(-30,0)(60,0){$\chi_{i,i}$, $\psi_{i,i+1}$}
  \end{picture}
\end{center}
\begin{align}
  \label{eq:fermions}
  \chi_{i,i} \qquad &\text{transforming as} \ (m, \bar n, 1) \\
  \psi_{i,i+1} \qquad &\text{transforming as} \ (1, n, \bar m)
\end{align}
where $i=1,\dots,N$ (and $i=0$ is periodically identified with $i=N$).

The field theory defined by this diagram is both anomaly and
asymptotically free for a wide range of $m$ and $n$.  At distances
short compared to both $1/\Lambda$ and $1/\Lambda_s$, the theory is
well-described by ($N$ copies of) four-dimensional weakly interacting
massless fermions and gauge bosons.

What does the theory look like at longer distances? In the limit where
$\Lambda_s \gg \Lambda$ the long distance behavior is also simple. At
energy scales near $\Lambda_s$ the $SU(m)$ gauge coupling is quite
weak, and may be treated perturbatively. At this scale each of the
$SU(n)$ groups become strong, causing the fermions to condense in
pairs: a non-zero expectation value forms for each pair of fermions
connected to a given strong gauge group:
\begin{equation}
  \label{eq:condense}
  \langle \chi_{i,i} \,\psi_{i,i+1} \rangle \sim 4 \pi
  f_s^3\,U_{i,i+1} \qquad i=1,\dots,N
\end{equation}
where $f_s \sim \Lambda_s/(4\pi)$ and $U_{i,i+1}$ is an $m\times m$
unitary matrix parameterizing the direction of the condensate. The
confining strong interactions also produce a spectrum of ``hadrons'',
analogues of ordinary glueballs and baryons, all with masses on the
order of $\Lambda_s\sim 4\pi f_s$.  Below the scale $\Lambda_s$ the
theory can be described as a $\Pi_1^N SU(m)$ gauge theory coupled to
$N$ non-linear sigma model fields, each transforming as
\begin{equation}
  U_{i,i+1} \to \mathbf{g}^{-1}_i(x) U_{i,i+1} \mathbf{g}_{i+1}(x)\ .
\end{equation}
We may use a diagram similar to the original moose to describe this
``condensed'' theory:
\begin{figure}[htb]
  \centering
  \scalebox{.65}{%
    \begin{picture}(0,325)(0,-144.8)
      \thicklines
      \put(-102.4,102.4){\rotatebox{45}{\usebox\cmoose}}
      \put(0,144.8){\usebox\cmoose}
      \put(102.4,102.4){\rotatebox{-45}{\usebox\cmoose}}
      \put(144.8,0){\rotatebox{-90}{\usebox\cmoose}}
      \put(102.4,-102.4){\rotatebox{-135}{\usebox\cmoose}}
      \put(0,-144.8){\rotatebox{-180}{\usebox\cmoose}}
      \put(-102.4,-102.4){\rotatebox{-225}{\usebox\cmoose}}
      \put(-144.8,0){\rotatebox{270}{%
          \makebox(0,0){\dottedline{10}(-50,0)(50,0)}}}
      \put(-60,144.8){\makebox(0,0){\bf G}}
      \put(60,144.8){\makebox(0,0){\bf G}}
      \put(60,-144.8){\makebox(0,0){\bf G}}
      \put(-60,-144.8){\makebox(0,0){\bf G}}
      \put(144.8,60){\makebox(0,0){\bf G}}
      \put(144.8,-60){\makebox(0,0){\bf G}}
      \put(-144.8,60){\makebox(0,0){\bf G}}
      \put(-144.8,-60){\makebox(0,0){\bf G}}
    \end{picture}
  }
  \caption{A condensed moose diagram}
  \label{fig:moose2}
\end{figure}
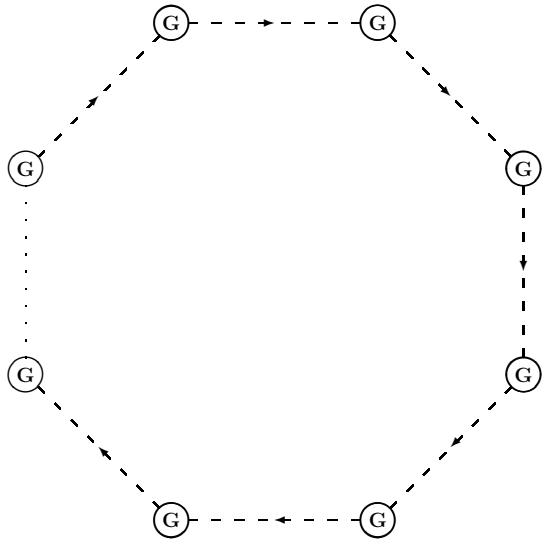

The low-energy effective action for this non-linear sigma model is
\begin{equation}
  \label{eq:lattice}
  S =\int\!\! d^4\!x\, \Biggl( -\frac{1}{2g^2} \sum_{j=1}^N
  \text{tr}\,  F_j^2 + f_s^2
  \sum_{j=1}^N \text{tr}\, \Bigl[ (D_\mu U_{j,j+1})^\dagger D^\mu
  U_{j,j+1}\Bigr] +\dotsb \Biggr)
\end{equation}
where the covariant derivative is $D_\mu U_{j,j+1} \equiv \partial_\mu
U_{j,j+1} -i A^j_\mu U_{j,j+1} + i U_{j,j+1} A^{j+1}_\mu$ and the dots
represent higher dimension operators that are irrelevant at low
energies. The action for the non-linear sigma model fields connects
the gauge fields at neighboring sites. In fact we
recognize~\eqref{eq:lattice} as simply a discretized action for a
five-dimensional gauge theory with gauge group $SU(m)$, where only the
fifth dimension has been latticized. The non-linear sigma model fields
are precisely the link variables of a lattice gauge theory, and the
condensed moose diagram is a picture of the fifth dimension! It is
remarkable that the moose diagram has transformed from a mnemonic for
the particle content of a four-dimensional gauge theory to a new
physical dimension of space at large distances.

The lattice spacing and the circumference of the fifth dimension are
\begin{equation}
  \label{eq:spacing}
  a=\frac{1}{gf_s}, \qquad R = N a\ .
\end{equation}
The five-dimensional gauge coupling is
\begin{equation}
  \label{eq:coupling}
  \frac{1}{g_5^2} = \frac{1}{a g^2} = \frac{f_s}{g}\ .
\end{equation}
We may eliminate any lingering doubt as to the five dimensional nature
of this theory by calculating the spectrum of the $N$ gauge
multiplets.  The fluctuations of the condensates $U_{i,i+1}$ higgs the
gauge group down to the diagonal subgroup.  The gauge boson mass
matrix is
\begin{equation}
  \label{eq:mmatrix}
  g^2 f_s^2  \begin{pmatrix}
    2      &     -1 &      0 &      0 &      0 &  \dots & -1 \\
    -1     &      2 &     -1 &      0 &      0 &  \dots & 0 \\
    0      &     -1 &      2 &     -1 &      0 &  \dots & 0 \\
    \vdots & \vdots & \ddots & \ddots & \ddots & \vdots & \vdots \\
    0      &  \dots &      0 &      0 &     -1 &      2 & -1 \\
    -1     &  \dots &      0 &      0 &      0 &     -1 & 2
  \end{pmatrix}
\end{equation}
This matrix is familiar from the physics of balls and springs, and its
eigenvalues are easily calculated~\cite[for example]{Georgi:waves},
yielding a mass spectrum labeled by an integer $k$ satisfying
$-N/2<k\leq N/2$,
\begin{equation}
  \label{eq:spectrum}
    M^2_k = 4 g^2 f_s^2 \sin^2\left(\frac{\pi k}{N}\right) \equiv
  \left(\frac{2}{a}\right)^2 
  \sin^2\left(\frac{p_5a}{2}\right) \ ,
\end{equation}
where $p_5\equiv 2\pi k/R$ is the discrete five-dimensional momentum.  The
corresponding eigenvectors are of the form $\psi^m \sim \exp(i\, m\,
p_5 a)$.  For $\abs{k} \ll N/2$ the masses become
\begin{equation}
  \label{eq:large}
  M_k \simeq \abs{p_k} = \frac{2\pi\abs{k}}{R}\ .
\end{equation}
This is precisely the Kaluza-Klein spectrum for a five-dimensional
gauge boson compactified on a circle of circumference $R$. The gauge coupling
of the diagonal subgroup is $g_4^2 = g^2/N$, and
using~\eqref{eq:coupling} this gives
\begin{equation}
  \label{eq:fcoupling}
   \frac{1}{g_4^2} = \frac{R}{g_5^2} \ ,
\end{equation}
the usual relation between the five-dimensional and four-dimensional
coupling constants.

\section{What is a fifth dimension?}
\label{sec:dimension}

For those familiar with lattice gauge theory, the appearance of the
lattice action~\eqref{eq:lattice} makes clear that, in every sense, a
true fifth dimension has appeared at large distances.\footnote{Perhaps
  not \emph{every} sense. We have not yet included gravitational
  interactions in our theory.}  Nevertheless it is productive to
examine how five-dimensional physics is reproduced.

The static potential between widely separated test charges is a common
measure of the dimensionality of space: the potential is $r^{2-d}$ in
$d$ spatial dimensions.  Consider two charges at sites $\ell$ and $j$
separated by a distance $r$ in the ordinary three dimensions. The
potential is a sum of Yukawa potentials from the KK modes, multiplied
by the couplings to each mode
\begin{equation}
  \label{eq:vsum}
    V_{j\ell}(r) \sim \frac{g^2}{N} \sum_k 
    e^{i(j-\ell) p_k a} \,\frac{e^{-M_k r}}{r} 
\end{equation}
For $ a\ll r \ll R$ the masses $M_k$ that contribute significantly
to~\eqref{eq:vsum} can be accurately approximated by \eqref{eq:large}
and the sum in~\eqref{eq:vsum} can be approximated by an integral,
\begin{equation}
  \label{eq:vint}
  V(r) \sim   \frac{g^2}{N} \int_{-\infty}^\infty dk\,
  e^{2 \pi ik a(j-\ell)/R} \,\frac{e^{-2\pi \abs{k} r/R}}{r} 
  \sim \frac{g_5^2}{a^2 (j-\ell)^2 + r^2} = \frac{g_5^2}{r_5^2 + r^2}
\end{equation}
where $r_5 = a(j-\ell)$ is the distance between the charges in the
fifth dimension. The inverse-square fall-off is a sure sign of a fifth
dimension, and is very different from the behavior of the potential
at distances much shorter than $a$ where the theory looks
four-dimensional. At distances much shorter than both $1/\Lambda$ and
$1/\Lambda_s$, the potential between particles charged under any one
of the gauge groups falls as $1/r$ (modulo an additional slow
variation due to the running of the gauge coupling).  The contrast is
even more striking for the potential between fermions charged under
different gauge groups, corresponding to test charges separated in the
fifth direction. At these short distances, this potential falls off
much more rapidly: the leading interaction between different sites
$(\ell,j)$ comes from a $2|\ell-j|+1$ loop diagram, and so falls off
exponentially with $|\ell-j|$ as $g^{4|\ell-j|+4}$.

Every other physical measurement performed at distances much larger
than $a$ but much smaller than $R$, will reveal a fifth dimension.
Since the fifth dimension emerged dynamically, rather than being put
in by hand, it is worthwhile to briefly address the question: what
\emph{is} a fifth dimension?  Mathematically, any set of ordered points
can be called a ``dimension'', but physically we need more. Particles
should be able to move in the extra dimension; that is, they should
carry labels, their co-ordinates in the fifth dimension, that change
as they move in the fifth dimension.  Furthermore, there should be a
physical notion of locality in the extra dimension. This translates
into the requirement of locality for the interactions in the theory.
Particles with the same labels have the largest interaction, while
particles with very different labels should interact only weakly.

These are the two defining properties of an extra dimension, and the
fifth dimension we have generated possesses both of them. The gauge
bosons propagate in the fifth dimension. Locality is a consequence of
the nearest neighbor coupling structure of our moose, enforced by our
choice of fermion content, gauge invariance and renormalizability.
These constraints have another interesting consequence.  There are
infinitely many possible latticizations of a fifth-dimension, each
with a different spectrum. We might then suspect that the
spectrum~\eqref{eq:spectrum} is easily modified. In fact, the
particular lattice action~\eqref{eq:lattice} and the corresponding
spectrum~\eqref{eq:spectrum} followed uniquely from our renormalizable
theory: the moose has made its choice.

\section{Lorentz invariance}
\label{sec:gun}

Extra dimensions may or may not be endowed with other properties as
well.  For instance they may be translationally invariant, or posses
the full higher-dimensional Lorentz symmetry. Whether or not these
additional properties arise in our constructions is a dynamical
question.  In the simple model we have presented, translational
invariance is manifest, and the full $SO(4,1)$ Lorentz invariance also
emerges at distances larger than $a$.

The lattice structure of the fifth dimension breaks five dimensional
Lorentz invariance.  For simplicity we will consider the limit $R\to
\infty$, where the theory appears five-dimensional at arbitrarily long
distances.  In this limit $p_5$ becomes a continuous variable, and the
dispersion relation for the five-dimensional gauge boson becomes
\begin{equation}
  E^2 = {\vec p}^{\,2}  + \left(\frac{2}{a}\right)^2
  \sin^2\left(\frac{p_5 a}{2}\right) \to {\vec p}^{\,2}  +   p_5^2
  \quad\text{as $a\to 0$.}
\end{equation}
When $a\to 0$ the five dimensional Lorentz invariance is automatically
restored. This might seem surprising because the fifth dimension is
apparently quite different from the other three space dimensions. But
in the limit of tiny spacing the only possible difference is a scale
choice that we have eliminated to leading order in the weak coupling
by defining the lattice spacing~\eqref{eq:spacing}. Quantum effects
will produce small changes in~\eqref{eq:spacing}, but there is
\emph{a} definition of the lattice spacing that produces a Lorentz
invariant limit in the full quantum theory. The fact that Lorentz
invariance is automatic in this continuum limit is a consequence of
the simplicity of this construction. In more complicated models,
five-dimensional Lorentz invariance in the continuum limit may require
tuning of parameters.

The violations of five dimensional Lorentz invariance due to the
finite lattice size $a$ appear as a sequence of higher dimension
operators in the five dimensional theory suppressed by powers of
$p_5a$. We expect contributions of this size in any five dimensional
theory because the inverse lattice spacing $1/a$ plays the role of a
cut-off. The difference here, compared to a standard five dimensional
effective theory, is that the high energy theory above the cut-off
scale is well defined, but lacks five dimensional Lorentz invariance.
Thus the interactions suppressed by powers of the cut-off are
calculable, but some break the Lorentz symmetry. Above the cut-off
scale, there is no vestige of five dimensional Lorentz symmetry
remaining, because the theory is perfectly four-dimensional at short
distances.

\section{Phase structure}
\label{sec:phase}

The field theory associated with the diagram in
figure~\ref{fig:moose1} has an obvious symmetry between $\text{\bf G}$
and $\text{\bf G}_s$. It is clear that the discussion above for
$\Lambda_s\gg\Lambda$ can be repeated for $\Lambda\gg\Lambda_s$. In
this dual situation, the physics is described by the dual of the
condensed moose in figure~\ref{fig:moose2}.  The condensates in this
case are
\begin{equation}
  \label{eq:condensedual}
   \langle \psi_{i,i+1} \,\chi_{i+1,i+1} \rangle \sim 4\pi
  f^3\,V_{i,i+1}  \qquad i=1,\dots,N\ .
\end{equation}
Again the physics is five dimensionful for $a\ll r\ll R$, but it is a
different fifth dimension, dynamically generated by a different set of
interactions and with a different set of gauge bosons.

The transition from~\eqref{eq:condense} to~\eqref{eq:condensedual} is
theoretically fascinating, but somewhat puzzling, and we will not
discuss it in detail here.  But it is important to understand the
approach to the transition because it bears on the possibly
phenomenologically relevant question of how large the five dimensional
gauge coupling can be.  For example the heaviest of the KK modes has a
mass of order $g(\Lambda_s) f_s$, parametrically lighter than the
scale where $\text{\bf G}_s$ gets strong, $\Lambda_s$. How similar can
we make these scales? Can we increase $g$ to the region of strong
coupling as well?  

For simplicity, let us take $N$ to infinity so that the physics
appears five-dimensional at arbitrarily long distances.  What happens
as we change the ratio of $\Lambda$ to $\Lambda_s$?  For
$\Lambda\ll\Lambda_s$, where the analysis of section~\ref{sec:moose}
applies, the residual gauge interactions at distances large compared
to $a$ are very weak. In the five dimensional language, this is
obvious because the gauge coupling is dimensional, $g_5^2=g^2a$, and
its effects at distances of order $\ell$ are suppressed by powers of
$g_5^2/\ell$. In the four dimensional language, one might worry that
there is something wrong with this argument at distances large
compared to $1/\Lambda$, but such worry is groundless. The weak gauge
group is higgsed by the condensate~\eqref{eq:condense} down to a
residual gauge group with coupling of order $g^2/N$ and thus becomes
arbitrarily weak as we take $N\to\infty$.

What happens as we increase $\Lambda/\Lambda_s$? The gauge coupling
$g_5^2=g^2a$ increases, but its effects remain tiny at large
distances. We know that at some point as $\Lambda\to\Lambda_s$, an
ecological disaster will occur, dramatically changing the nature of
the long distance physics. But it is reasonable to suppose that the
cataclysm will happen abruptly at some point
$\Lambda\approx\Lambda_s$, where both gauge couplings are strong.  The
only signal at large distances of impending doom is that as
$\Lambda\to\Lambda_s$, $g_5^2=g^2a$ gets large compared to $a$. This
signals the imminent breakdown of the effective theory because
dimensional couplings in an effective theory must not be large
compared to the appropriate power of the cut-off.  Even though the
tree level interactions are still weak at long distances, the theory
is losing control of its quantum corrections, a warning that anarchy
is about to be loosed upon the world.

\section{Other completions}
\label{sec:other}

The fifth dimension has appeared in the condensed moose because the
non-linear sigma model fields allow the gauge field to ``hop'' from
one site to the next. Since we could have obtained this directly as a
latticization of the five-dimensional gauge theory, we might ask why
we need the original moose model at all.  The reason is that
latticization in the fifth dimension does not cut-off divergences from
large four-momenta: the four-dimensional non-linear sigma model
of~\eqref{eq:lattice} is non-renormalizable, becoming strongly coupled
at a scale $\sim 4\pi f_s$. That is, \emph{this} theory requires a UV
completion.  But this is familiar problem, with familiar solutions.
The moose model we have constructed provides a UV completion in the
same way that QCD completes the theory of pions. However purely
perturbative completions are also possible. For example we could
replace the non-linear sigma model with a renormalizable, linear sigma
model; we replace each unitary field $U_{i,i+1}$ with a charged scalar
field $\phi_{i,i+1}$.  The action for this sigma-model will include a
quartic potential for these scalars. If this potential produces vacuum
expectation values for all the scalars at a scale $f_s$, this model is
indistinguishable from our moose model at low energies. A fifth
dimension then appears just as before.

In the linear sigma-moose, it is easy to include other degrees of
freedom at the sites.  Including appropriate couplings to the link
variables will allow these fields to hop in the extra dimension as
well. For example Yukawa couplings to fermions at the sites will
produce hopping. Since the strength of this hopping term is unrelated
to the gauge coupling, the fermions and gauge bosons propagate with
different maximal speeds in the extra dimension. The resulting theory
is five-dimensional, but without five-dimensional Lorentz invariance,
even at large distances, although we can always tune the couplings to
recover Lorentz invariance at long distance.

Although the linear sigma model example is renormalizable, the natural
value for the vacuum expectation values of the scalar fields is the UV
cut-off of the four-dimensional theory. We can avoid this standard
problem of fundamental scalars in a standard perturbative way: by
using supersymmetry.  For simplicity we consider an $N=1, \text{\bf
  G}=SU(2)$ SUSY gauge version of our condensed moose, although
extensions to larger gauge groups are straightforward.  The arrows in
this case are meaningless because the fundamental and anti-fundamental
of $SU(2)$ are the same. The line connecting $i$ to $i+1$ denotes a
bi-fundamental chiral superfield $\phi_i$, that we can think of as a
$2 \times 2$ matrix.  In addition to the gauge interactions, the
theory has a superpotential
\begin{equation}
  \label{eq:superpot}
  W = \lambda \sum_i S_i (\det \phi_i - \mu^2)
\end{equation}
where $S_i$ are gauge singlet chiral fields, and $\mu$ is a mass
scale. The theory is asymptotically free as long as the ratio
$\lambda/g$ is not too large.  The superpotential forces spontaneous
symmetry breaking. Writing $\phi_i = (\mu + A_i)\exp(\Sigma_i^a
\sigma^a)$, the superpotential pairs up $A_i$ and $S_i$ with a mass
$\sim \lambda \mu$, while the $\Sigma_i$ contain the massless
Goldstone bosons (together with their superpartners). At low energies
we are left with a latticization of the five-dimensional $N=1$ $SU(2)$
gauge theory.  In components, we have $SU(2)$ gauge bosons, together
with a Dirac fermion and a real scalar in the adjoint representation.
Note that in this theory no tuning of parameters is required to obtain
five-dimensional Lorentz invariance.  Supersymmetry guarantees that
the gauge bosons, fermions and scalars propagate in the
fifth-dimension with the same maximum velocity, and full
five-dimensional Lorentz invariance is recovered at long distances.

\section{Conclusions and speculations}
\label{sec:conclusions}

We have constructed a fifth dimension dynamically in a
four-dimensional \emph{renormalizable} gauge theory. At long distances
the physics is that of a compactified five-dimensional gauge theory
with dimensionful couplings, that by itself would be
non-renormalizable.  This construction is easily extended to produce
several extra dimensions.

We can now investigate higher-dimensional physics in a well-defined
setting. Questions involving energies higher than the na\"\i{}ve
five-dimensional cut-off are straightforward in this context.
Higher-dimensional phenomena, such as power-law running, localization
of gauge fields and chiral fermions, orbifold compactification and
supersymmetry breaking, to name a few, have straightforward
constructions in our formalism~\cite{forthcoming}.

The insight provided by our technique can work in both directions.
Just as constructing extra dimensions in a renormalizable setting
illuminates higher-dimensional physics, so too the physics of extra
dimensions may suggest new phenomena in the context of purely four
dimensional models that have no extra-dimensional interpretation.
This has led to a novel approach for stabilizing the electroweak
scale~\cite{preparation}.

How does gravity fit in? The simplest possibility is to add
four-dimensional gravity to our four-dimensional field theories.
While the non-gravitational physics appears five-dimensional, gravity
remains purely four-dimensional.  Constructing extra dimensions in
this way frees us from many of the na\"\i{}ve constraints of
higher-dimensional model building. In particular the absence of
gravity in the fifth-dimension eliminates many of the defects of
non-standard gravity at high energies. For example radius
stabilization is no longer an issue---there is no dynamical radius to
stabilize! Rather the size of the extra dimension is set by the fixed
parameters of the four-dimensional theory.  As another example, the
cosmology of extra dimensions is often troublesome. But in our
construction, the Universe at temperatures above the na\"\i{}ve
five-dimensional cut-off is described by a completely standard
four-dimensional FRW cosmology.  Without gravity the shape of the
extra dimensions is not constrained by Einstein's equations.  In fact
the extra dimensions we have constructed may not have any simple
manifold interpretation at all (consider a ``figure-8'')!  It is also
interesting to attempt to generate full five-dimensional gravity
through a similar mechanism.  This requires degrees of freedom that
link the four-dimensional geometry at each site.

It is tempting to imagine that some or all of the three ordinary
spatial dimensions may be generated dynamically. There is no obstacle
in principle to constructing moose models in $2+1$ dimensions that
generate a fourth dimension for non-gravitational fields. However a
mechanism for obtaining four-dimensional gravity is essential.

The dynamical generation of extra dimensions within four-dimensional
field theories allows exploration of higher-dimensional physics in a
familiar context. Conversely insights from extra dimensions may be
applied directly to purely four-dimensional models.  Our construction
serves as a link from extra dimensions to a new world of ideas.

\section*{Acknowledgements}

H.G. is supported in part by the National Science Foundation under
grant number NSF-PHY/98-02709. A.G.C. is supported in part by the
Department of Energy under grant number \#DE-FG02-91ER-40676.  N.A-H.
is supported in part by the Department of Energy. under Contracts
DE-AC03-76SF00098, the National Science Foundation under grant
PHY-95-14797, the Alfred P. Sloan foundation, and the David and
Lucille Packard Foundation.


\providecommand{\href}[2]{#2}\begingroup\raggedright
\endgroup

\end{document}